\documentclass{article}
\usepackage{amsthm}
\usepackage{amsmath}
\PassOptionsToPackage{numbers, compress}{natbib}

\usepackage[preprint]{neurips_2023}

\usepackage[utf8]{inputenc} 
\usepackage[T1]{fontenc}    
\usepackage{hyperref}       
\usepackage{url}            
\usepackage{booktabs}       
\usepackage{amsfonts}       
\usepackage{nicefrac}       
\usepackage{microtype}      
\usepackage{xcolor}         
\usepackage{graphicx}
\usepackage{subcaption}
\usepackage[export]{adjustbox}

\title{Delta Score: Improving the Binding Assessment of Structure-Based Drug Design Methods}

\author{
Minsi Ren\textsuperscript{1}, 
Bowen Gao\textsuperscript{2}\,, 
Bo Qiang\textsuperscript{3}\,, 
\textbf{Yanyan Lan\textsuperscript{2}}\thanks{Correspondence to \texttt{lanyanyan@air.tsinghua.edu.cn}}  \\
\textsuperscript{1}Institute of Automation, Chinese Academy of Sciences \\
\textsuperscript{2}Institute for AI Industry Research (AIR), Tsinghua University \\
\textsuperscript{3}Department of Pharmaceutical Science, Peking University 
}

\begin{document}

\maketitle

\begin{abstract}
Structure-based drug design (SBDD) stands at the forefront of drug discovery, focusing on developing molecules that target specific binding pockets. Recent advances in this area have witnessed the adoption of deep generative models, modeling SBDD as a conditional generation task where the target structure serves as context. Despite previous claims that generated ligands outperform their respective ground truth counterparts in terms of docking score evaluation, our analysis reveals that these perceived performance improvements are attributed to inherent biases within the scoring systems themselves, rather than an accurate assessment of the ligands' binding affinity. To address this issue, we introduce the delta score, a new evaluation metric 
emphasizing docking scores that prioritize specificity. Our experiments reveal that molecules produced by current deep generative models significantly lag behind ground truth reference ligands when assessed with the delta score. This novel metric not only complements existing benchmarks but also provides a pivotal direction for subsequent research in the domain. 
\end{abstract}


\section{Introduction}

In the field of drug discovery, the development of novel small molecules that could form stable binding complexes with a specific disease-related target, known as structure-based drug design (SBDD) tasks, is of paramount importance. With the availability of an increasing amount of structural data, such as the PDBbind\cite{pdbbind} and Crossdocked dataset\citep{crossdock}, many deep generative models~\cite{ar, pocket2mol, shape2mol} have been proposed to address SBDD by formulating it as a target-based conditional generation task, resulting in remarkable progress.


Specifically, the assessment of these models has predominantly focused on docking scores between generated molecules and designated targets, using docking software such as Vina \citep{AutoDockVina}. Despite the claims made by state-of-the-art models that a majority of the ligands they generate outperform the docking score of ground truth ligands in test sets~\cite{targetdiff}, it raises 
the questions: Has the issue of 3D molecule generation been conclusively resolved? Do these scores align with real-world biological needs?


To address these inquiries, we conducted an experiment on the CrossDocked2020 dataset to achieve a fair comparison between various deep generative models. Our reproduced outcomes have exhibited evident enhancements of these deep generative models compared to the ground-truth, at least for a certain proportion of targets, which are consistent with previous claims. However, surprisingly, we discovered that a basic random sampling approach from Zinc also produced superior outcomes when compared to the ground-truth. This exceptional result prompts us to reevaluate the chosen evaluation metric.

Our analysis show the shortcomings of using “docking score” as the primary evaluation metric: the overall averaged score could be artificially inflated by characterizing certain biases, without truly capturing the matching degree between the molecules and the target. Therefore, we propose a complementary metric named "delta score", which allows us to distinguish specificity in binding to the target by subtracting the average matching degree, thereby better accommodating the requirements of applications in structure-based drug molecule design.

\section{Related Work}\label{related_work}


With the emergence of geometric models~\citep{satorras2022en, geiger2022e3nn}, the field of Structure-Based Drug Design (SBDD) has shifted towards 3D neural networks for encoding protein structures and decoding 3D molecule conformations, representing real-world 3D interactions. Various methods have been proposed, including voxel-based methods~\citep{masuda2020generating}, auto-regressive models~\citep{luo20213d, pocket2mol, graphbp}, and diffusion models~\citep{targetdiff, guan2023decompdiff}. Most of these works used Vina score~\cite{AutoDockVina} for binding affinity evaluation, which is a typical docking software to predict the interactions between a small molecule and a protein target, similar to Glide \citep{Glide} and Gold \citep{gold}. However, there is limited discussion on the suitability of this widely used docking metric in assessment of SBDD methods.

\section{Experimental Analysis w.r.t. Docking Score}
We first conduct experiments on CrossDocked2020 dataset~\citep{crossdock} to compare different deep generative models, including auto-regressive model\citep{ar} (denoted as AR), Shape2mol\citep{shape2mol}, Pocket2mol\citep{pocket2mol} and Targetdiff\citep{targetdiff}. In addition, we randomly sample small molecules for each target from Zinc dataset \citep{irwin2005zinc}, to form a baseline named Random\_Zinc method. The data preprocessing and splitting are all following ~\citet{ar} to ensure a fair comparison. For each method, we generate 100 molecules for each target pocket in the test set and calculate the affinity score using Glide instead of Vina, since Glide has demonstrated superior performance in predicting both accurate conformation and binding affinity~\citep{wang2016comprehensive}. It is important to note that the Vina evaluation results display a high degree of similarity, therefore the conclusions drawn can be generalized across different docking scores.

\subsection{Experimental Results}
From the results in Table \ref{glidescore}, we find that: 1) there is relatively little difference in the averaged docking scores generated by different models, including ground truth and Random\_Zinc; 2) Targetdiff consistently outperforms ground truth; 3) each method is capable of generating molecules that outperform ground truth performance on different targets, for example, Targetdiff outperforms ground truth on nearly half targets and even Random\_Zinc has the ability to outperform ground truth on 22.4\% targets. Evidently, these findings have sparked concerns regarding the reliability of the docking metrics utilized.




\begin{table}[htbp]
    \centering
    \caption{Glide Docking Scores}
    \label{glidescore}
    \begin{tabular}{c |c c c }
        \hline
        \noalign{\vskip 2pt}
        \textbf{Dataset} & \multicolumn{3}{c}{CrossDocked 2020}\\
        \textbf{Methods} & \textbf{Mean of mean ($\downarrow$)} & 
        \textbf{Median of mean ($\downarrow$)} &
        \textbf{Better than GT ($\uparrow$)} \\
        \noalign{\vskip 2pt}
        \hline
        \noalign{\vskip 2pt}
        Ground Truth & -6.367 & -6.581 &  -\\
        \noalign{\vskip 2pt}
        \hline
        \noalign{\vskip 2pt}
        AR & -5.833 & -5.666 & 0.359\\
        \noalign{\vskip 2pt}
        \hline
        \noalign{\vskip 2pt}
        Pocket2mol & -6.282 & -6.170 & 0.382\\
        \noalign{\vskip 2pt}
        \hline
        \noalign{\vskip 2pt}
        Shape2mol & -5.631 & -5.663 & 0.265\\
        \noalign{\vskip 2pt}
        \hline
        \noalign{\vskip 2pt}
        Targetdiff & -6.670 & -6.742 & 0.489\\
        \noalign{\vskip 2pt}
        \hline
        \noalign{\vskip 2pt}
        Random\_Zinc & -5.543 & -5.564 & 0.224\\
        \noalign{\vskip 2pt}
        \hline
    \end{tabular}
\end{table}

\subsection{Case Study}
Upon further analysis of the small molecules generated by Targetdiff, particularly those exhibiting favorable docking scores, we have discovered some intriguing properties:
\begin{itemize}
\item[$\bullet$] These molecules typically exhibit highly intricate cyclic structures, implying their extremely low likelihood of existing in reality and the significant challenges to synthesize. 
\item[$\bullet$] These molecules typically have promising docking scores against most of pockets in the test set, rather than only their specific target, which means they may be pan-assay interference compounds (PAINS) or have poor specificity.
\end{itemize}

\begin{figure}[htbp]
    \centering
\begin{subfigure}[b]{0.48\textwidth}
        \captionsetup{singlelinecheck=false,justification=centering}
        \includegraphics[width=\textwidth,valign=t]{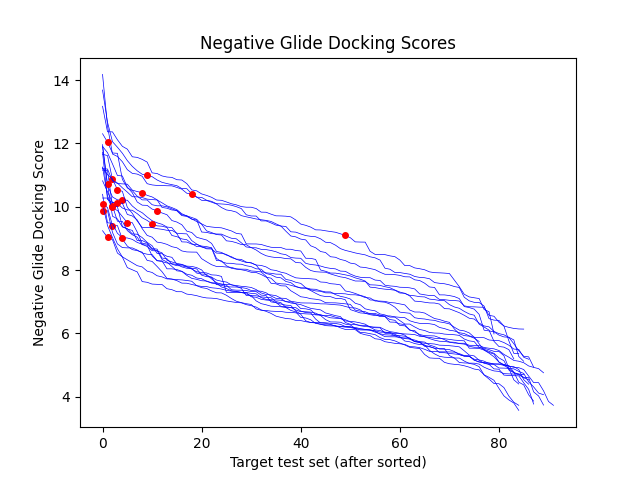}
        \caption{Docking Scores against targets in the test set after sorted (multiplied by -1)}
        \label{fig:left}
\end{subfigure}
\hspace{0.05cm}
\begin{subfigure}[b]{0.48\textwidth}
        \captionsetup{singlelinecheck=false,justification=centering}
        \includegraphics[width=\textwidth,valign=t]{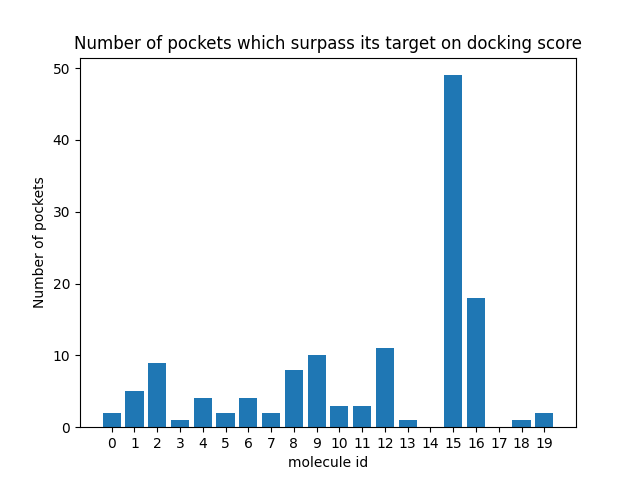}
        \caption{The number of pockets which surpass true target on docking score}
        \label{fig:right}
\end{subfigure}
    \caption{In the left image, each line represents the sorted docking score of a small molecule against all pockets in the test set. The highlighted red represents its true target. The right image displays the number of test pockets for each small molecule in which the docking score is higher than their respective true target.}
    \label{fig:case study 1}
\end{figure}

More specifically, we randomly select 20 pockets in the test set and generate 100 small molecules for each pocket. We select the one with the best docking score on each pocket, totaling 20 small molecules. All these small molecules have a docking score of over -9 on their true targets. We cross dock them with all the pockets in the test set, the results are shown as Figure \ref{fig:case study 1}.
We notice that out of the 20 small molecules, only 2 have a top 1 docking score for their true targets. Furthermore, we have observed that for almost a quarter of the molecules, the docking scores are higher for over 10 other pockets as compared to their true targets. For a more detailed demonstration, we choose to display one generated molecule with a glide docking score of up to -10.111 against its target pocket in Appendix B.

\subsection{Analysis}
Based on previous experimental results and case studies, we have concluded that current docking scores have limitations in accurately evaluating the binding relationships between generated molecules and targets, leading to false positive issues.

According to \citep{lyu2023modeling,ferreira2015molecular,mysinger2010rapid}, docking software generally employs force field models to assess the interaction energy between molecules and receptors. However, these models are often empirical and trained on known structures and properties, resulting in inherent limitations. They may fail to accurately capture all molecular features and types of interactions, leading to biases. One consequence of this bias is that docking software may assign high scores to small molecules with some certain specific structures, such as PAINS, even if they are actually false positives. 

While docking scores can offer insights into the binding affinity between small molecules and target proteins, they may not provide a complete assessment of selectivity. It is crucial to consider the off-target effects, which arise when a drug molecule interacts with unintended targets, leading to potential adverse reactions and impacting the overall therapeutic outcome \citep{harrison2016phase, wong2019estimation,lin2019off}.

To develop an metric that can effectively counteract bias, a straightforward yet effective way is to consider the difference in scores between positive pairs and negative pairs rather than solely focusing on the score of one pair of molecule and receptor. Based on this, we introduce delta score in Section 4.1. By incorporating the variation in affinity scores across different targets, this metric can also assess the specific binding capacity of a small molecule to its target.

\section{Delta Score}

\subsection{Definition}
Suppose the test set contains $n$ target pockets $ {p_1,p_2...p_i...,p_n}$, for each pocket $p_i$ the model generates $m$ molecules $x_{i1},x_{i2}...x_{ij}...,x_{im}$.
Using docking software mentioned in Section 3.1, we calculate the affinity score between a molecule and a pocket: $S(x_{ij}, p_i)$. We define the binding ability of small molecules generated by the model for target $p_i$ to target $p_k$ as: 
\begin{equation}
    \text{BindingAbility}_{ik} = 
    \mathbb{E}_{j \in (1,m)}[S(x_{ij}, p_i)]
\end{equation}
In order to strengthen specificity and reduce the effect of PAINS fragments, we defined a novel metric, the delta score for SBDD. The metric is defined as follow:
\begin{equation}\label{delta_socre}
\begin{split}
    \text{DeltaScore}(p_i) = \text{BindingAbility}_{ii} - \text{BindingAbility}_{ik,k\neq i}
    \\= \mathbb{E}_{j \in (1,m)}[S(x_{ij}, p_i)] - \mathbb{E}_{j \in (1,m)}[S(x_{ij}, p_k)_{ k\neq i}] 
\end{split}
\end{equation}
We have also proposed a sampling technique in Appendix A, to improve the computational efficiency of approximating this expectation.

\bibliographystyle{plainnat}
\subsection{Experimental Results w.r.t. Delta Score}
\begin{table}[htbp]
    \centering
    \caption{Results on CrossDocked dataset with delta score. \\ Best results are \underline{underlined}.}
    \label{DeltaScores}
    \begin{tabular}{c|c c}
        \hline
        \noalign{\vskip 2pt}
        \textbf{Dataset} & \multicolumn{2}{c}{CrossDocked 2020}\\
        \textbf{Methods} & \textbf{mean of mean ($\downarrow$)} & \textbf{median of mean ($\downarrow$)} \\
        \noalign{\vskip 2pt}
        \hline
        \noalign{\vskip 2pt}
        Ground Truth & \underline{-0.810} & \underline{-1.062} \\
        \noalign{\vskip 2pt}
        \hline
        \noalign{\vskip 2pt}
        AR & -0.535 & -0.296 \\
        \noalign{\vskip 2pt}
        \hline
        \noalign{\vskip 2pt}
        Pocket2mol & -0.309 & -0.231 \\
        \noalign{\vskip 2pt}
        \hline
        \noalign{\vskip 2pt}
        Shape2mol & 0.052 & -0.072 \\
        \noalign{\vskip 2pt}
        \hline
        \noalign{\vskip 2pt}
        Targetdiff & -0.382 & -0.505 \\
        \noalign{\vskip 2pt}
                \hline
        \noalign{\vskip 2pt}
        Random\_Zinc & -0.019 & -0.018 \\
        \noalign{\vskip 2pt}
        \hline
    \end{tabular}
\end{table}
We have conducted a reevaluation of state-of-the-art deep generative models on CrossDocked2020 with a focus on delta score. As indicated in Table~\ref{DeltaScores}, we have observed the correction of all exceptional results in Table~\ref{glidescore}. Firstly, there are significant differences among different methods, with ground truth being the best and random being the worst. This is reasonable because Random\_Zinc could be considered as an unconditional random sampling method, its target aware binding performance is expected to close to zero. This reaffirms our belief that the delta score can incisively evaluate the conditional components of molecules, pinpointing those elements that genuinely and effectively engage with the target structure. Secondly, even the best generative models, e.g.~AR and Targetdiff, shows significant differences from the ground truth, indicating that there is still a considerable research and development space in this direction.

\section{Conclusion}
While contemporary deep generative techniques have elevated the docking score, our analysis indicates that this enhancement predominantly pertains to the non-conditional aspects. Such improvements, we deduce, can be attributed to the methods learning biases that might mislead docking software, occasionally leading to the generation of anomalous molecular structures. We envision this pioneering metric—delta score—as a valuable addition to the current set of benchmarks. By providing deeper insights, we hope it will pave the way for more informed advancements in the realm of Structure-Based Drug Design.

\bibliography{refs}

\begin{thebibliography}{23}
\providecommand{\natexlab}[1]{#1}
\providecommand{\url}[1]{\texttt{#1}}
\expandafter\ifx\csname urlstyle\endcsname\relax
  \providecommand{\doi}[1]{doi: #1}\else
  \providecommand{\doi}{doi: \begingroup \urlstyle{rm}\Url}\fi

\bibitem[Ferreira et~al.(2015)Ferreira, Dos~Santos, Oliva, and Andricopulo]{ferreira2015molecular}
Leonardo~G Ferreira, Ricardo~N Dos~Santos, Glaucius Oliva, and Adriano~D Andricopulo.
\newblock Molecular docking and structure-based drug design strategies.
\newblock \emph{Molecules}, 20\penalty0 (7):\penalty0 13384--13421, 2015.

\bibitem[Francoeur et~al.(2020)Francoeur, Masuda, Sunseri, Jia, Iovanisci, Snyder, and Koes]{crossdock}
Paul~G Francoeur, Tomohide Masuda, Jocelyn Sunseri, Andrew Jia, Richard~B Iovanisci, Ian Snyder, and David~R Koes.
\newblock Three-dimensional convolutional neural networks and a cross-docked data set for structure-based drug design.
\newblock \emph{Journal of chemical information and modeling}, 60\penalty0 (9):\penalty0 4200--4215, 2020.

\bibitem[Geiger and Smidt(2022)]{geiger2022e3nn}
Mario Geiger and Tess Smidt.
\newblock e3nn: Euclidean neural networks, 2022.

\bibitem[Guan et~al.(2023{\natexlab{a}})Guan, Qian, Peng, Su, Peng, and Ma]{targetdiff}
Jiaqi Guan, Wesley~Wei Qian, Xingang Peng, Yufeng Su, Jian Peng, and Jianzhu Ma.
\newblock 3d equivariant diffusion for target-aware molecule generation and affinity prediction.
\newblock \emph{arXiv preprint arXiv:2303.03543}, 2023{\natexlab{a}}.

\bibitem[Guan et~al.(2023{\natexlab{b}})Guan, Zhou, Yang, Bao, Peng, Ma, Liu, Wang, and Gu]{guan2023decompdiff}
Jiaqi Guan, Xiangxin Zhou, Yuwei Yang, Yu~Bao, Jian Peng, Jianzhu Ma, Qiang Liu, Liang Wang, and Quanquan Gu.
\newblock Decompdiff: Diffusion models with decomposed priors for structure-based drug design.
\newblock 2023{\natexlab{b}}.

\bibitem[Halgren et~al.(2004)Halgren, Murphy, Friesner, Beard, Frye, Pollard, and Banks]{Glide}
Thomas~A. Halgren, Robert~B. Murphy, Richard~A. Friesner, Hege~S. Beard, Leah~L. Frye, W.~Thomas Pollard, and Jay~L. Banks.
\newblock Glide: a new approach for rapid, accurate docking and scoring. 2. enrichment factors in database screening.
\newblock \emph{Journal of Medicinal Chemistry}, 47\penalty0 (7):\penalty0 1750--1759, 2004.

\bibitem[Harrison(2016)]{harrison2016phase}
Richard~K Harrison.
\newblock Phase ii and phase iii failures: 2013--2015.
\newblock \emph{Nat Rev Drug Discov}, 15\penalty0 (12):\penalty0 817--818, 2016.

\bibitem[Irwin and Shoichet(2005)]{irwin2005zinc}
John~J Irwin and Brian~K Shoichet.
\newblock Zinc- a free database of commercially available compounds for virtual screening.
\newblock \emph{Journal of chemical information and modeling}, 45\penalty0 (1):\penalty0 177--182, 2005.

\bibitem[Lin et~al.(2019)Lin, Giuliano, Palladino, John, Abramowicz, Yuan, Sausville, Lukow, Liu, Chait, et~al.]{lin2019off}
A~Lin, CJ~Giuliano, A~Palladino, KM~John, C~Abramowicz, ML~Yuan, EL~Sausville, DA~Lukow, L~Liu, AR~Chait, et~al.
\newblock Off-target toxicity is a common mechanism of action of cancer drugs undergoing clinical trials. sci. transl. med. 11: eaaw8412, 2019.

\bibitem[Liu et~al.(2022)Liu, Luo, Uchino, Maruhashi, and Ji]{graphbp}
Meng Liu, Youzhi Luo, Kanji Uchino, Koji Maruhashi, and Shuiwang Ji.
\newblock Generating 3d molecules for target protein binding.
\newblock \emph{arXiv preprint arXiv:2204.09410}, 2022.

\bibitem[Long et~al.(2022)Long, Zhou, Dai, and Zhou]{shape2mol}
Siyu Long, Yi~Zhou, Xinyu Dai, and Hao Zhou.
\newblock Zero-shot 3d drug design by sketching and generating.
\newblock \emph{Advances in Neural Information Processing Systems}, 35:\penalty0 23894--23907, 2022.

\bibitem[Luo et~al.(2021{\natexlab{a}})Luo, Guan, Ma, and Peng]{ar}
Shitong Luo, Jiaqi Guan, Jianzhu Ma, and Jian Peng.
\newblock A 3d generative model for structure-based drug design.
\newblock \emph{Advances in Neural Information Processing Systems}, 34:\penalty0 6229--6239, 2021{\natexlab{a}}.

\bibitem[Luo et~al.(2021{\natexlab{b}})Luo, Guan, Ma, and Peng]{luo20213d}
Shitong Luo, Jiaqi Guan, Jianzhu Ma, and Jian Peng.
\newblock A 3d generative model for structure-based drug design.
\newblock \emph{Advances in Neural Information Processing Systems}, 34:\penalty0 6229--6239, 2021{\natexlab{b}}.

\bibitem[Lyu et~al.(2023)Lyu, Irwin, and Shoichet]{lyu2023modeling}
Jiankun Lyu, John~J Irwin, and Brian~K Shoichet.
\newblock Modeling the expansion of virtual screening libraries.
\newblock \emph{Nature Chemical Biology}, pages 1--7, 2023.

\bibitem[Masuda et~al.(2020)Masuda, Ragoza, and Koes]{masuda2020generating}
Tomohide Masuda, Matthew Ragoza, and David~Ryan Koes.
\newblock Generating 3d molecular structures conditional on a receptor binding site with deep generative models, 2020.

\bibitem[Mysinger and Shoichet(2010)]{mysinger2010rapid}
Michael~M Mysinger and Brian~K Shoichet.
\newblock Rapid context-dependent ligand desolvation in molecular docking.
\newblock \emph{Journal of chemical information and modeling}, 50\penalty0 (9):\penalty0 1561--1573, 2010.

\bibitem[Peng et~al.(2022)Peng, Luo, Guan, Xie, Peng, and Ma]{pocket2mol}
Xingang Peng, Shitong Luo, Jiaqi Guan, Qi~Xie, Jian Peng, and Jianzhu Ma.
\newblock Pocket2mol: Efficient molecular sampling based on 3d protein pockets.
\newblock In \emph{International Conference on Machine Learning}, pages 17644--17655. PMLR, 2022.

\bibitem[Satorras et~al.(2022)Satorras, Hoogeboom, and Welling]{satorras2022en}
Victor~Garcia Satorras, Emiel Hoogeboom, and Max Welling.
\newblock E(n) equivariant graph neural networks, 2022.

\bibitem[Trott et~al.(2009)Trott, Oleg, Olson, Arthur, and J.]{AutoDockVina}
Trott, Oleg, Olson, Arthur, and J.
\newblock Autodock vina: Improving the speed and accuracy of docking with a new scoring function, efficient optimization, and multithreading.
\newblock \emph{J. Comput. Chem.}, 31\penalty0 (2):\penalty0 NA--NA, 2009.

\bibitem[Verdonk et~al.(2003)Verdonk, Cole, Hartshorn, Murray, and Taylor]{gold}
Marcel~L Verdonk, Jason~C Cole, Michael~J Hartshorn, Christopher~W Murray, and Richard~D Taylor.
\newblock Improved protein--ligand docking using gold.
\newblock \emph{Proteins: Structure, Function, and Bioinformatics}, 52\penalty0 (4):\penalty0 609--623, 2003.

\bibitem[Wang et~al.(2005)Wang, Fang, Lu, Yang, and Wang]{pdbbind}
Renxiao Wang, Xueliang Fang, Yipin Lu, Chao-Yie Yang, and Shaomeng Wang.
\newblock The pdbbind database: methodologies and updates.
\newblock \emph{Journal of medicinal chemistry}, 48\penalty0 (12):\penalty0 4111--4119, 2005.

\bibitem[Wang et~al.(2016)Wang, Sun, Yao, Li, Xu, Li, Tian, and Hou]{wang2016comprehensive}
Zhe Wang, Huiyong Sun, Xiaojun Yao, Dan Li, Lei Xu, Youyong Li, Sheng Tian, and Tingjun Hou.
\newblock Comprehensive evaluation of ten docking programs on a diverse set of protein--ligand complexes: the prediction accuracy of sampling power and scoring power.
\newblock \emph{Physical Chemistry Chemical Physics}, 18\penalty0 (18):\penalty0 12964--12975, 2016.

\bibitem[Wong et~al.(2019)Wong, Siah, and Lo]{wong2019estimation}
Chi~Heem Wong, Kien~Wei Siah, and Andrew~W Lo.
\newblock Estimation of clinical trial success rates and related parameters.
\newblock \emph{Biostatistics}, 20\penalty0 (2):\penalty0 273--286, 2019.

\end{thebibliography}
\appendix
\section{Delta Score}
To mitigate off-target effects, our new metric is designed to emphasize molecules with a specific binding affinity to the target protein pockets, as opposed to binding to multiple similar pockets without selectivity. As a result, we anticipate that all generated molecules should exhibit a negative score, as follows:

\begin{equation}
\begin{split}
    \text{BindingAbility}_{ii} - min_{k \in (1,n), k\neq i}\text{BindingAbility}_{ik}= 
    \\\mathbb{E}_{j \in (1,m)}[S(x_{ij}, p_i)] - min_{k \in (1,n), k\neq i} \mathbb{E}_{j \in (1,m)}[S(x_{ij}, p_k)] < 0
\end{split}
\end{equation}

Notice that to calculate $min_{k \in (1,n), k\neq i} \mathbb{E}_{j \in (1,m)}[S(x_{ij}, p_k)]$, we need to perform docking for molecules with all possible pockets which is of quadratic complexity. To address this challenge, instead of docking molecules with every pockets, we randomly sample $\tilde{n}$ pockets outside of the pocket $p_i$: $\{p_{1'}...p_{\tilde{n}'}\} \subset \{ p_1, p_2,...,p_n \}\setminus\{p_i\} $ and define Delta Score for $p_i$ as:

\begin{equation}
    \text{DeltaScore}_{\tilde{n}}(p_i) = \mathbb{E}_{j \in (1,m)}[S(x_{ij}, p_i)] - min_{k \in (1',\tilde{n}')} \mathbb{E}_{j \in (1,m)}[S(x_{ij}, p_k)] 
\end{equation}
When $\tilde{n} = n-1$, the Delta Score is defined as equivalent to Eq.~\ref{delta_socre}. In order to save computing resources, we set $\tilde{n} = 1$ in which case the delta score becomes:

\begin{equation}\label{delta_socre1}
    \text{DeltaScore}_1(p_i) = \mathbb{E}_{j \in (1,m)}[S(x_{ij}, p_i)] - \mathbb{E}_{j \in (1,m)}[S(x_{ij}, p_k)_{ k\neq i}] 
\end{equation}

It can be readily demonstrated that, in terms of statistical significance, this is equivalent to the difference between the model's binding affinity for a specific target and the model's average binding affinity for other targets:

\begin{equation}
    \mathbb{E}_i[\text{DeltaScore}_1(p_i)] \equiv 
    \mathbb{E}_i[ \text{BindingAbility}_{ii} - \frac{1}{n-1}\sum_{k \in (1,n), k \neq i}{ \text{BindingAbility}_{ik}}] 
\end{equation}

\begin{proof}[Proof]
According to equation \ref{delta_socre},

\begin{align}
&\mathbb{E}_{i\in(1,n)}[\text{DeletaScore}_1(p_i)]  \\ 
&=\mathbb{E}_{i\in(1,n)}\mathbb{E}_{j \in (1,m)}[S(x_{ij}, p_i)] - \mathbb{E}_{i\in(1,n)}\mathbb{E}_{j \in (1,m)}[S(x_{ij}, p_k)_{ k\neq i}] \\
&=\mathbb{E}_i[\text{BindingAbility}_{ii}] - \frac{1}{n}\sum_{i,k}^{i\neq k} \mathbb{E}_j[S(x_{ij}, p_k)] \\  
&\equiv \mathbb{E}_i[\text{BindingAbility}_{ii}] - \frac{1}{n}\sum_{i}\frac{1}{n-1}\sum_k^{k\neq i} \mathbb{E}_j[S(x_{ij}, p_k)] \\ &= \mathbb{E}_i[\text{BindingAbility}_{ii}] - \mathbb{E}_i[\frac{1}{n-1}\sum_{k \in (1,n), k \neq i} \text{BindingAbility}_{ik}] \\
&= \mathbb{E}_i[\text{BindingAbility}_{ii} - \frac{1}{n-1}\sum_{k \in (1,n), k \neq i} \text{BindingAbility}_{ik}] 
\end{align}

\end{proof}

\section{Case Study}
\begin{figure}[htbp]
    \centering
\begin{subfigure}[b]{0.45\textwidth}
        \captionsetup{singlelinecheck=false,justification=centering}
        \includegraphics[width=\textwidth,valign=t]{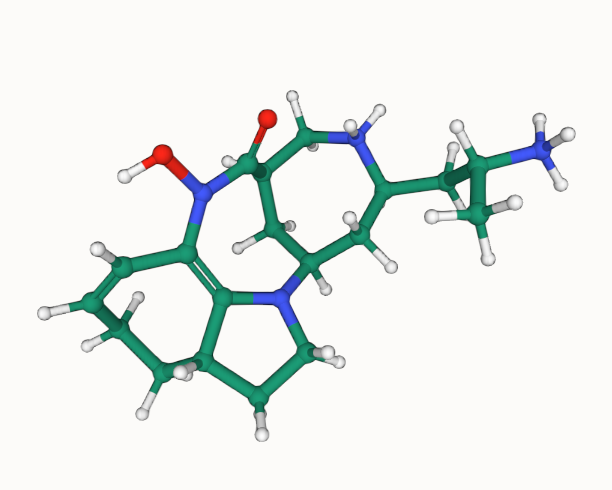}
        \caption{2D structural diagram of the molecule}
        \label{fig:left}
\end{subfigure}
\hspace{1cm}
\begin{subfigure}[b]{0.45\textwidth}
        \captionsetup{singlelinecheck=false,justification=centering}
        \includegraphics[width=\textwidth,valign=t]{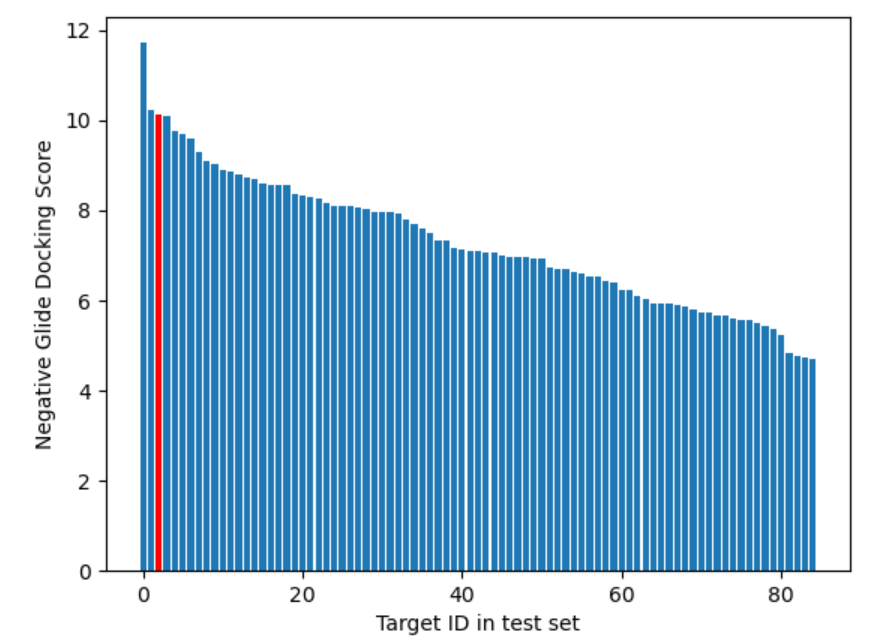}
        \caption{Docking Scores against targets in the test set (multiplied by -1)}
        \label{fig:right}
\end{subfigure}
    \caption{ One molecule generated by Targetdiff Model against 4lfu\_A\_rec.pdb pocket. The left image shows its 2D structure. The right image shows its docking scores against pockets in the test set (after sorted). The column highlighted in red indicates its actual target. }
    \label{fig:case study 2}
\end{figure}

We choose one generated molecule against 4lfu\_A\_rec.pdb pocket for a more detailed display. It is questionable and has many issues as depicted in Figure \ref{fig:case study 2}. On the one hand, the molecular skeleton of the compound is complicated, consisting of multiple non-aromatic ring structures. The chemical reactions that could directly obtain this molecular skeleton are limited. The lipid rings exhibit poor water solubility and is easy to be metabolized. The N-hydroxamide is not common in drugs. On the other hand, we observe that it exhibits fairly good docking scores for the majority of the pockets in the test set, with an average of -7.02. Two pockets even surpass its actual target receptor on docking score, which indicates this molecule is highly likely to be a pan-assay interference compound or prone to off-target effect. Similar phenomenon occurs on almost all generated molecules with good docking scores, which indicates that only using docking score metric is far from enough to evaluate the effectiveness of current generative model.

\end{document}